\RequirePackage{fix-cm}
\documentclass[smallextended]{svjour3}       
\smartqed  
\usepackage{graphicx}
\usepackage{amsmath}
\usepackage{mathabx}
\usepackage{comment}

\usepackage{color}
\usepackage{array,bold-extra,tikz}
\usetikzlibrary{automata,arrows,shapes,topaths,trees,backgrounds,shadows}
\usepgfmodule{matrix}
\usepackage{mathtools}
\usepackage{times}

\usepackage{float}

\usepackage{natbib}
\setcitestyle{aysep={}} 

\usepackage{url}
\usepackage{tabularx}
\usepackage[utf8]{inputenc}
\usepackage{epigraph}
\usepackage{epigraph}

\usepackage{etoolbox}

\makeatletter
\patchcmd{\epigraph}{\@epitext{#1}}{\itshape\@epitext{#1}}{}{}
\makeatother

\usepackage{csquotes}

\renewcommand{\phi}{\varphi}

\newcommand{\BOX}[2]{[#1_{#2}]}
\newcommand{\dia}[2]{\langle #1_{#2}\rangle}

\newcommand{\B}[1]{\BOX{\leq}{#1}}
\newcommand{\dB}[1]{\dia{\leq}{#1}}

\newcommand{\Agt}{\mathcal{I}}

\def\L{\mathcal{L}}

\def\M{\mathcal{M}}

\def\L{\mathcal{L}}

\def\M{\mathcal{M}}
\def\A{\mathcal{A}}

\def\Agt{\mathcal{I}}

\newcommand{\HM}[1]{\textcolor{red}{[Huimin: #1]}}

\newcommand{\uncong}{\mathord{\cong}}
\newcommand{\unll}{\mathord{>}}

\begin{document}

\title{Dynamic Logic of Legal Competences~\thanks{Huimin Dong is supported by the National Social Science Fund of China (No. 20CZX051, No. 17ZDA026) and the National Science Centre of Poland
(No. UMO-2017/26/M/HS1/01092). Both authors are supported by the PIOTR research project (No. RO 4548/4-1). The authors would like to thank Alessandra Marra, Paul McNamara, Piotr Kulicki, and the anonymous reviewers for their valuable remarks and comments. 
	}
}


\author{Huimin Dong \and Olivier Roy 
}


\institute{Huimin Dong \at
              Department of Philosophy (Zhuhai), Sun Yat-sen University \\
              \email{Huimin.Dong@xixilogic.org}           
           \and
           Olivier Roy \at
           Institut f\"{u}r Philosophie, Universit\"{a}t Bayreuth\\
           \email{Olivier.Roy@uni-bayreuth.de}
}

\date{Received: date / Accepted: date}

\maketitle

\begin{abstract}
  We propose a new formalization of legal competences, and in
particular for the Hohfeldian categories of power and immunity,
through a deontic reinterpretation of dynamic epistemic logic. We
argue that this logic explicitly captures the norm-changing character of
legal competences while providing a sophisticated reduction of
the latter to static normative positions. The logic is completely
axiomatizable, and we apply it to a concrete case in German contract
law to illustrate that it can capture the distinction between legal
ability and legal permissibility.

\end{abstract}

\vspace{15pt}

The Hohfeldian typology of rights~\citep{hohfeld1913some} distinguishes
what one might call \emph{static} and \emph{dynamic} rights. Static
rights encompass claims and privileges, as well as their
respective correlatives duties and no-claim. They have also been called ``normative positions''~\citep{sergot2001computational}. On the dynamic side
one finds \emph{legal competences}\footnote{What we
	call here static and dynamic rights have been labelled in various
	ways in the formal literature. Kanger called static rights the
	``type of the states of affairs'' and dynamic ones the ``type of
	influence''~\citep{kanger1972law}. Makinson instead used the
	``deontic family'' and the ``legally capacitative family'' for
	static and dynamic rights,
	respectively~\citep{makinson1986formal}. Bentham, von Wright and Hart
	on the other hand used ``legal validity'' and ``norm-creating
	action''~\citep{lindahl1977position}, while
	Lindahl~\citeyearpar{lindahl1977position} called it ``the range of
	action.'' } : power and immunity and their correlative
liability and no-power.\footnote{The terminology used by Hohfeld might not fully reflect how terms like ``power'', ``immunity'' or ``liability'' have been used in all the English-speaking literature. The work following~\citep{kanger1972law} has, however, in important parts used that terminology. In other to situate our contribution better in that tradition, we use it here as well. We thank one of the anonymous reviewers of the JLLI for pointing us to the importance of making this caveat.} See Figure~\ref{legrig}. In this paper we will be mainly focusing on legal competences. 

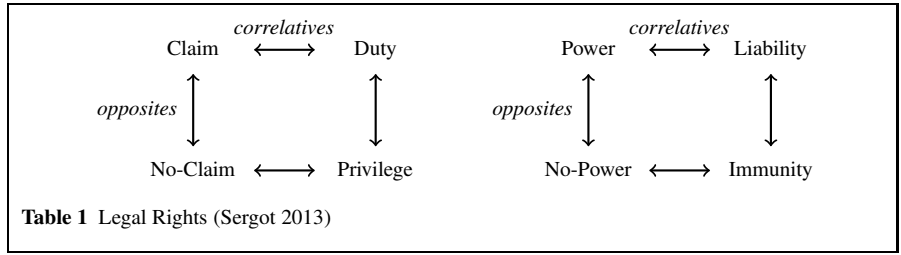
\begin{table}[htbp]
	\fbox{
		\parbox{.96\textwidth}
		{\centering
			\begin{tabular}{lcr}
				\begin{tikzpicture}[scale=.8]
				
				\tikzstyle{rel}=[draw,very thick]
				\tikzstyle{relA}=[draw,black,thick, <->]
				\tikzstyle{relB}=[rel,dotted,black,<->]
				\tikzstyle{relC}=[rel,dashed,blue,->]

				\tikzstyle{cell}=[draw,thick,rounded corners,dashed]
				\tikzstyle{cell_1}=[cell,solid]
				\tikzstyle{cell_2}=[cell,dotted]
				\tikzstyle{cell_3}=[cell,dashdotted]

				\coordinate (cellnw) at (-1,1);
				\coordinate (cellne) at (1,1);
				\coordinate (cellse) at (1,-1);
				\coordinate (cellsw) at (-1,-1);
				
				\tikzstyle{state}=[circle, fill, inner sep=2pt]
				\tikzstyle{empty}=[]

				\node[empty, label = below:{Claim}] (c) at (-2,2) {};
				\node[empty, label = below:{Duty}] (d) at (1,2) {};
				\node[empty, label = below:{No-Claim}] (nc) at (-2,0) {};
				\node[empty, label = below:{Privilege}] (pr) at (1,0) {};

				\node[empty, label = above:{{\it correlatives}}] (co) at (-.5,1.6) {};
				\node[empty, label = left:{{\it opposites}}] (op) at (-2,.6) {};

				\path[relA] (-1,1.6) -- (0,1.6);
				\path[relA] (-1,-.4) -- (0,-.4);
				\path[relA] (-2,1.2) -- (-2,0);
				\path[relA] (1,1.2) -- (1,0);
				
				\end{tikzpicture}
				
				& &
				
				\begin{tikzpicture}[scale=.8]
				
				\tikzstyle{rel}=[draw,very thick]
				\tikzstyle{relA}=[draw,black,thick, <->]
				\tikzstyle{relB}=[rel,dotted,black,<->]
				\tikzstyle{relC}=[rel,dashed,blue,->]

				\tikzstyle{cell}=[draw,thick,rounded corners,dashed]
				\tikzstyle{cell_1}=[cell,solid]
				\tikzstyle{cell_2}=[cell,dotted]
				\tikzstyle{cell_3}=[cell,dashdotted]

				\coordinate (cellnw) at (-1,1);
				\coordinate (cellne) at (1,1);
				\coordinate (cellse) at (1,-1);
				\coordinate (cellsw) at (-1,-1);
				
				\tikzstyle{state}=[circle, fill, inner sep=2pt]
				\tikzstyle{empty}=[]

				\node[empty, label = below:{Power}] (po) at (3,2) {};
				\node[empty, label = below:{Liability}] (l) at (6,2) {};
				\node[empty, label = below:{No-Power}] (np) at (3,0) {};
				\node[empty, label = below:{Immunity}] (im) at (6,0) {};
				
				\node[empty, label = above:{{\it correlatives}}] (co) at (4.5,1.6) {};
				\node[empty, label = left:{{\it opposites}}] (op) at (3,.6) {};
				
				\path[relA] (4,1.6) -- (5,1.6);
				\path[relA] (4,-.4) -- (5,-.4);
				\path[relA] (3,1.2) -- (3,0);
				\path[relA] (6,1.2) -- (6,0);

				\end{tikzpicture} \\
			\end{tabular}
			\caption{Legal Rights~\citep{marek2011positions}} \label{legrig}
	}}
\end{table}

Although logical approaches to legal competences are scarcer than
for static normative positions, existing theories can be divided into
two broad families. The first formalizes power and immunity
as (legal) permissibility, or absence thereof, to see to it that a
certain normative position obtains~\citep{kanger1966rights,
	lindahl1977position}. Lindahl~\citeyearpar{lindahl1977position}, for instance, defines
$j$'s power to make it the case that $i$ ought to see to it that
$\phi$ as a permission that $j$ sees to it that $i$ ought to see to it that $\phi$:
\[P(Do_jO(Do_i\phi))\] We call such an approach reductive because it
takes power and immunity as definable in the language of obligations,
permissions, and actions, where claims and privileges are also
defined. Non-reductive approaches, on the other hand, view power and
immunity as position-changing
actions~\citep{makinson1986formal,jones1996formal,van2016modelling} or normative
conditionals~\citep{governatori2005temporalised} that are not reducible
to static normative positions. 

Both of these families have assets and drawbacks.  Reductive approaches
come with a rich logical theory of the relationship between static
normative positions and legal competences, with the latter inheriting
its logic from the former. Choosing to define power as above,
however, obfuscates the changing potential of legal competence by
reducing it to permissibility, a simple static legal relation. This
potential for change was arguably crucial for Hohfeld who defined power as
the ability to ``\emph{change} legal
relations''~\citep[p.44-45]{hohfeld1913some}.  The formalization above,
in particular, furthermore conflates legal ability (\emph{rechtliches
	K\"{o}nnen}) with legal permissibility (\emph{rechtliches
	D\"{u}rfen}), although these two concepts are
distinct~\citep{makinson1986formal}. Non-reductive approaches, on the
other hand, do better justice to the dynamic character of legal
competences by taking norm-changing actions or conditionals as first
class citizens in the logic. This allows to distinguish legal ability
and legal permissibility. The cost of this, however, is a logic of legal competences, which is, at least at the outset, independent from the logic of the static normative positions.

The dynamic logic that we present in this paper provides a
middle ground between these two types of approach. It is reductive,
and as such comes with a rich set of principles of interaction between
static and dynamic rights. It does so, however, while retaining both
the dynamic character of legal competences and the distinction between
legal ability and legal permissibility.

What we propose is a deontic re-interpretation of dynamic \emph{epistemic} logic~\citep{van2007dynamic,van2011logical}, and should be mainly seen as a contribution to that field. Indeed, the
reader familiar with it will recognize both the modeling methodology
and the axiomatization that we present here. What we show is that this
formalism also yields interesting insights when interpreted in
deontic terms, especially for theories of legal competence.\footnote{The results presented in this paper build on and extend the work reported
	in~\citep{dong2017dynamic}. The present paper uses a more general
	condition for static conditional obligations, presents an improved
	version of the definitions of power and immunity, and a more
	in-depth conceptual discussion of the key points.}

Our formalization of legal competences draws also from~\cite{markovichSL}'s recent proposal for the analysis
of powers. Like ours, her approach uses tools from dynamic epistemic
logic, namely public announcement operators, to explicate Hohfeldian
powers. Our underlying logic of static rights is furthermore taken
directly from her work~\citep{markovich2016agents}, as it constitutes
one of the most sophisticated accounts of \emph{directed} rights currently available. Our analysis, however, generalizes and complements hers in
at least two ways. First, the theory we propose here uses general
deontic action operators, for which public announcements are special
cases. As we will see, this allows us to analyze ``softer'',
i.e. defeasible kinds of deontic actions. Second, our analysis of
static rights is based on a logic with conditional obligation operators,
in which standard operators of standard deontic logic are easily definable. Finally, our
analysis addresses two points which are either implicit or only
mentioned in passing in Markovich's work, namely the problem of vacuous powers when formalized using DEL tools, and the distinction
between legal permissibility and legal ability.

Another interesting recent contribution to which our proposal relates to is~\citet{van2016modelling}, which proposes a formalization of the Holfeldian categories using tools from propositional dynamic logic~\citep{harel2000dynamic}. The mathematical relationship between the two proposals needs to be investigated further. The following points should, however, be highlighted at the outset. On the one hand, \citeauthor{van2016modelling}'s formalization allows for a more fine-grained description of complex actions than the language we use here, by capturing explicitly, for instance, concurrent or sequential actions as well as non-deterministic choices. They also explicitly include an epistemic component to address the question of knowledge-based obligations, which we leave out here. On the other hand they make two important simplifying assumptions. They first restrict their analysis of legal competences to abilities to change simple claim rights about atomic propositions. They furthermore explicitly leave out cases where a third party, for instance the judiciary, can change the legal relations obtaining between two other agents. As we will see below we consider these kinds of cases to be of primary importance for a logical model of legal competence. 

The rest of the paper is structured as follows. In Section~\ref{sec:4.1} we present the underlying model of static rights, and then move on to dynamic modalities
and legal competences in Section~\ref{sec:4.2}. We show how the two can be put
together to capture the four Hohfeldian basic types of right, and we present
a complete axiomatization. We then apply it to a concrete case in
the German civil code in order to show that legal ability and legal
permissibility can be naturally distinguished in this logic.

\section{Static Rights}
\label{sec:4.1}
This paper is not about static rights, but rather about legal
competences. Therefore, as to the former we stay as close as possible to the
mainstream theory of normative positions stemming
from the proposals developed by~\cite{kanger1972law,lindahl1977position,makinson1986formal} and
surveyed by~\cite{marek2011positions}, but also incorporating
directionality as proposed by~\cite{markovich2016agents,markovichSL}. This theory,
at least in its non-directional form, has been thoroughly studied when it comes to
claim rights and privileges. It has, of course, well-known drawbacks,
which we inherit as well. The reader uncomfortable with this modeling
choice should keep in mind that the dynamic tools
that we deploy later are to a large extent modular. They may be
used with different static theories. We point to some possible
alternatives along the way.

The only difference between the mainstream logic for static rights,
including the work of~\cite{markovich2016agents,markovichSL}, and the one we use
here is that ours contains conditional duties and permissions. This is
a technical decision. Conditional duties allow for a simpler
integration with the dynamic part. In its deontic version these
conditional dynamic rights rest on the dyadic deontic logic developed
by~\cite{benthem2014priority}, but goes back at least
to~\cite{hansson1969analysis}.

\subsection{Language}
On the surface, the language we use differs from mainstream
normative positions in that it contains a family of Kripke modalities on an
underlying preference order, along with the universal modality and the usual
``seeing to it that'' modality.\footnote{There have been a number of
	different proposals for defining ``seeing to it that''. One
	popular family of approaches uses either one~\citep{chellas1980modal}
	or a pair of unary
	modalities~~\citep{kanger1972law,lindahl1977position,makinson1986formal}. All
	of them satisfy the {\it T} axiom as well as the {\it E} rule for
	substitution of logical equivalence. More recent approaches, as
	for instance the so-called ``Chellas STIT'', use a normal, {\bf S5}
	modality~\citep{chellas1992time,horty2000agency}.  } Again, we use this
language instead of the classical deontic one for technical
reasons. It facilitates the axiomatization of the dynamic
modalities. It is well known, however, that this language is
expressive enough to define conditional obligations and
permissions~\citep{boutilier1994toward,van2005preference}. We
will come back to this at the end of the section. The logic of
conditional obligation can be completely axiomatized, both in the
current language~\citep{van2007dynamic} or by taking only conditional
obligations as
primitive~\citep{baltag2008qualitative,parent2014maximality}.

\begin{definition} \label{def:lansta} Let $Prop$ be a countable set of
	atomic propositions and $i,j$ be elements of a given finite set $\Agt$ of agents. The language $\mathcal{L}$ is defined as follows:
	$$
	\phi := p  \mid \neg\phi \mid \phi \wedge \phi  \mid [\leq_{i\to j}]\phi \mid U\phi \mid Do_i\phi
	$$ 
	where $p \in Prop$. 
\end{definition}
\noindent We write $\langle \leq_{i\to j}\rangle \phi$ for
$\neg[\leq_{i\to j}]\neg\phi$, and $E\phi$ for $\neg U \neg \phi$. A
formula $U\phi$ is read as ``it is necessary that $\phi$.''
$Do_i \phi$ indicates a {\it non-deontic} or {\it
	ontic}~\citep{van2008semantic} action of agent $i$, and should be
read in the usual sense of ``$i$ sees to it that $\phi$.''

The modality $[\leq_{i\to j}]$ will be interpreted on a comparative ideality relation. Intuitively, in its undirected version, a formula $[\leq] \phi$ would say that $\phi$ is true in all the worlds at least as ideal as the current one.  In its directed version, the modality $[\leq_{i\to j}] \phi$ should be read as ``$\varphi$ is true in the worlds that are, \emph{as far as the obligations of $i$ towards $j$ are concerned}, at least as ideal as the current one." 

This directionality is important. Hohfeldian rights, both static and dynamic, are indeed ``directed''~\citep{sartor2005legal} or, in Makinson's words, ``resolutely relational''~\citep{makinson1986formal}. They specify both
their subjects and their addressees. The subjects are the agents who
hold the right, the addressees are those who have the correlative
duty. The standard formalization of static rights follows the
Kangerian tradition and leaves this directionality implicit. Indeed,
with the notable exception of some early work~\citep{kanger1966rights,herrestad1995obligations}, in this tradition neither the
subject $i$ appears in the formalization of claim rights, nor the
addressee $j$ in the formalization of privileges. Instead, when
writing $O(Do_j\phi)$ for ``$i$ has a claim against $j$ regarding
$\phi$'', it is assumed that $\phi$ is something that concerns $i$ in
a relevant manner. This modeling choice was already in place
in Kanger's work~\citep{kanger1972law}, but see a
recent overview by~\cite{marek2011positions}. Our language for static rights follows a recent
proposal by~\cite{markovich2016agents,markovichSL}, in which
this directionality is made explicit in the form of a family of
directed obligation operators of the form $O_{i \to j}$. In the
present language, as we explain below, both the conditional and the
unconditional forms of these operators are definable using the
$[\leq_{i\to j}]$ modality. There are of course alternative proposals
for capturing directed rights,
notably~\cite{governatori2005temporalised}'s work. We leave a full comparison
for future work.

\subsection{Semantics and Deontic Operators}
The semantics for the static part is constituted by so-called preferential
models~\citep{vanBenthem2010,benthem2014priority}, augmented with one
binary relation for each $Do_i$ operator. We do not assume that the
preference ordering is connected nor conversely well-founded. This
will give rise to slight differences from,
e.g.~\cite{vanBenthem2010,benthem2014priority}'s work.

\begin{definition}\label{epdox-model}  
	Let $Prop$ and $\Agt$ be as above.   A {\it preference-action model} $\M$ is a tuple
	$\langle W, \{\leq_{i\to j}, \sim_i\}_{i, j\in \Agt}, V\rangle$ where:
	\begin{itemize}
		\item $W$ is a non-empty set of states
		\item $\leq_{i\to j}$ is a reflexive and transitive relation on $W$
		\item for each $i\in\Agt$, $\sim_i$ is an equivalence relation on $W$
		\item $V: Prop \to \mathcal{P}(W)$ is a valuation function
	\end{itemize}
\end{definition}

The respective preference relations are interpreted in terms of comparative ideality or rather, in the present context, of comparative \emph{legal} ideality. In its undirected version, $w \leq u$ would be read as saying that $u$ is, from a legal point of view, at least as ideal as $w$. This comparative legal ideality is now restricted to the legal relation of $i$ towards $j$. So the interpretation of $w \leq_{i\to j} u$ becomes ``$u$ is, from a legal point of view, at least as ideal as $w$, as far as $i$'s obligations towards $j$ are concerned." 

Preference-action frames are preference-action models minus the
valuation. We assume that the relations $\sim_i$ are equivalence
relations to simplify the treatment of static rights and thus to put
the emphasis on our dynamic extension. This assumption could
be removed. Sentences of $\mathcal{L}$ are interpreted in
preference-action models. 
\begin{definition}
	The truth conditions for sentence $\phi \in \mathcal{\L}$ are defined as follows:
	\begin{displaymath}
	\begin{array}{lcl}
	\M, w \models p & \text{ iff } & w \in V(p)\\
	\M, w \models \neg \phi  & \text{ iff } &  \M, w \not\models \phi\\
	\M, w \models \phi \wedge \psi & \text{ iff } &  \M, w \models \phi \text{ and } \M, w \models \psi\\
	\M, w \models [\leq_{i\to j}] \phi  & \text{ iff } &  \M, w' \models \phi \text{ for all } w' \geq_{i\to j} w \\
	\M, w \models U \phi  & \text{ iff } & \M, w' \models \phi \text{ for all } w' \in W\\
	\M, w \models Do_i \phi  & \text{ iff } &  \M, w' \models \phi \text{ for all } w' \sim_i w
	\end{array}
	\end{displaymath}
\end{definition}
\noindent The validity of the models and frames, and the classes thereof, is defined as
usual. We define $||\phi ||$, the truth set of $\phi$, as
$\{w : \M, w \models\phi\}$.

As mentioned, directed conditional obligation, understood in terms of ``truth
in all the most ideal worlds'' is definable in this
language. In fact this language can define a stronger notion, that of \emph{conditional} directed obligation.\footnote{The complexity of the formula below is due to the fact that there might not be states that are maximal according to the relation  $\geq_{i \to j}$, i.e. states that have no other states strictly above them. When such states are guaranteed to exist, for instance  finite models, this definition reduces to the more familiar definition in terms of truth in all maximal states.} This definability argument is standard~\citep{burgess1981quick,boutilier1994toward,van2005preference}.
Indeed, let $O_{i \to j}(\psi/\phi)$ be defined as follows:
\begin{center}
	$\M, w \models O_{i\to j}(\psi/\phi)$ \\
	iff \\
	$\forall v \geq_{i \to j} w [\M, v \models \phi \Rightarrow \allowbreak \exists u \geq_{i\to j} v ( \M, u \models \phi \,\&\, \forall s \geq_{i\to j} u (\M, s \models \phi \Rightarrow \M, s \models \psi))]$
\end{center} 
\noindent  This directed conditional obligation operator is then definable as: 
\[O_{i\to j}(\psi/\phi) \equiv_{df} \B{i \to j}(\phi \rightarrow \dB{i\to
	j}(\phi \wedge \B{i\to j}(\phi \rightarrow \psi)))\] 
	The definition essentially states the truth condition of the conditional obligation. Recall that the latter says, in finite models, that all most ideal $\phi$-world, as far as $i$'s obligations towards $j$ are concerned, are also $\psi$-worlds. In the general case it says that for every $\phi$-world $v$ that is at least as ideal as the current one, there is a (possibly different) $\phi$-world $v'$ that is at least as ideal as $v$, from which any weakly more ideal world makes $\phi \rightarrow \psi$ true.

 In Hohfeldian terminology, given that duties correspond to a correlative claim, the formula $O_{i\to j}(\psi/\phi)$ can also be read as ``Given $\phi$, $j$ has a claim against $i$ regarding $\psi$."  Unconditional directed obligations $O_{i\to j}\phi$ are defined as
$O_{i\to j}(\phi /\top)$, while permission is ``weak permissions'',
i.e. $P_{i\to j}(\phi/\psi)$ iff $\neg O_{i\to j} (\neg \phi/\psi)$. A routine argument
shows that all these modalities are normal. They furthermore satisfy
the following qualified form of the $D$ axiom.
$$\dB{i\to j}\phi \rightarrow (O_{i\to j}(\psi/\phi) \rightarrow \neg O_{i\to j}(\neg \psi/\phi)) $$
It states that conditional directed obligations with the same conditions are
consistent unless their condition itself is not satisfiable in any world accessible from the current one.%

\subsection{Claims Rights and Privileges}

With this in hand we have the machinery required to define claims and privileges. We do so, again, using the directed Kangerian approach put forward
by~\citet{markovichSL}, which also contains an in-depth discussion of
this modeling of static rights in view of Hohfeld's original proposal
and the broader legal theory context:
\begin{itemize}
	\item Given $\psi$, agent $i$ has a claim against $j$ regarding $\phi$: $O_{j \to i}(Do_j\phi/\psi)$
	\item Given $\psi$, agent $i$ has a privilege against $j$ regarding
	$\phi$: $\neg O_{i \to j}(Do_i \neg \phi/\psi)$ or, equivalently,
	$P_{i \to j}(\neg Do_i \neg \phi/\psi)$
\end{itemize}

\subsection{Claims Rights and Privileges: An Example}

\begin{figure}[htbp]
\centering
	\begin{tikzpicture}[->,>=stealth',shorten >=.5pt,shorten <=.8pt,
	auto,node distance=.3cm,semithick, scale=.8]
	\tikzstyle{every state}=[draw=none,text=black]
	

	\node[state,label=left:{$\neg d, p, \neg f$}] at (0,0) (1) {{\small $\textcolor{black}{w_1}$}};
	\node[state,label=left:{$d, p , \neg f$}] at (0,2) (2) {{\small $w_2$}}; 
	\node[state,label=right:{$d, \neg p,\neg f$}] at (4,2) (3) {{\small $w_3$}};
	\node[state,label=right:{$\neg d, \neg p, \neg f$}] at (4, 0) (4) {{\small $w_4$}};
	
	\path[draw,dashed,thick] (-0.5,1.5) rectangle (0.5,2.5);
	
	\path[draw,dashed,thick] (-0.5,-.5) rectangle (0.5,0.5);
	
	\path[draw,dashed,thick] (3.5,1.5) rectangle (4.5,2.5);
	
	\path[draw,dashed,thick] (3.5,-.5) rectangle (4.5,.5);
	
	\path (2) edge[<-, draw=black] (1); 
	\path (3) edge[<->, draw=black] (4); 
	\path (2) edge[<->, draw=black] (3); 
	\path (1) edge[->, draw=black] (4);

	\end{tikzpicture}
	\caption{A static model of Ivy's example. All arrows represent the relation $\leq_{i \to c}$.}\label{fig:4.1} 
\end{figure}
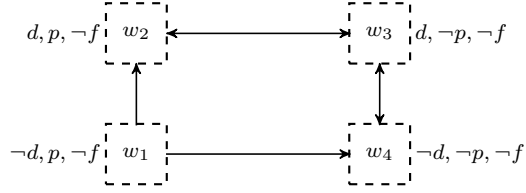

Here is an example to illustrate these notions, to which we will
return later.  Ivy ($i$) can park her car in a space that requires
the display of a parking permit in her windshield for monitoring by a city
official. Call $d$ the fact that the permit is displayed, and $p$ the
fact that the car is parked. The city ($c$) has a claim right against
Ivy to display the permit given that she has parked. In our setting this is
represented by the fact that the state where both $d$ and $p$ hold is
strictly better, with respect to $i$ towards $c$, than the state 
where the car is parked but the permit is not displayed ($p, \neg
d$). On the other hand, given that she hasn’t parked, the city has no
such right. Rather, Ivy has the freedom to display the permit or not.
This is illustrated by the double arrow between the two states to the
right of Figure~\ref{fig:4.1}. Suppose the current state is $w_1$, where Ivy has parked her car without a permit. Now, the
city has a conditional claim against Ivy to display the permit, $O_{i\to c}(Do_{i}d / p)$, but, on the other hand, Ivy has the
privilege both to display and not to display the permit conditional on not
parking her car, $\neg O_{i \to c}(Do_{i}d/\neg p)$ and
$\neg O_{i \to c}(\neg Do_{i}d/\neg p)$. Note, furthermore, that Ivy
has no \emph{unconditional} obligation towards the city to display the permit or
to park her car.

Parking without displaying the permit can of course have legal
consequences for Ivy, for instance that it is obligatory that she pays
a fine ($f$). The crucial point of this example, however, is that this
only happens if a  parking officer
with the legal power to, first, witness
that Ivy is parked without a permit and, second, issue a parking
ticket. Absent these two deontic actions, or more precisely this
combination of deontic observation and action, the city has no claim
against Ivy regarding the payment of a fine. This is illustrated here
by $f$ being false everywhere in the model. 

\section{Legal Competences} \label{sec:4.2} We now turn to our
proposal for modeling legal competences. It essentially follows the
so-called ``event models'' methodology developed
by~\cite{baltag2008qualitative} for epistemic modalities. The key
idea, in this epistemic environment, is to model the structure of a
particular learning event using the same tools as for an agent's
static information, that is Kripke models. The result of
updating one's knowledge or belief in the light of new information is
then computed using some form of restricted product of these
models. See the textbook~\citep{van2007dynamic} for details.

Transposed into our deontic context, the proposal is to model
explicitly the structure of deontic actions or legal competences using
what we call \emph{deontic action models}.

\subsection{Deontic Action Models}\label{sec:4.2-1}
We start with the definition of deontic action models, which are our
main modeling tools for legal competences. 

\begin{definition} \label{event-model} A {\it deontic action model
		$\A_i$ for agent i} is a tuple
	$$\langle A, \{\leq_{j \to k}^{\A_i}\}_{j,k \in \mathcal{I}}, Pre, Post\rangle$$
	where:
	\begin{itemize}
		\item $A$ is a non-empty finite set of deontic actions.
		\item $\leq^{\A_i}_{j \to k}$ is a reflexive and transitive relation on $A$ 
		\item $Pre: A \to \mathcal{L}$ is a precondition function.
		\item $Post: A \to (Prop \to \mathcal{L})$ is a post-condition
		function that assigns to each action and each atomic proposition a formula in $\mathcal{L}$. We assume that, for all
		$a$, $Post(a)$ differs only, at most, in finitely many elements of
		$Prop$ from the identity function.
	\end{itemize}
\end{definition}
\noindent $A$ is the set of deontic actions or the bases for legal competences.
The reader should be careful, however, in equating deontic actions with legal competences. There will be typically no one-to-one correspondence between the elements of a deontic action model and the legal competences of a given agent. Often, the latter will be better
represented using \emph{sets} of actions in $A$, together with the relation $\leq^{\A_i}_{j \to k}$. Our running example, which
we present below, is a case in point. Deontic actions are thus components of legal competences, but not equal to them.

Typically, legal competences can only be enacted in specific
circumstances. A sale, for instance, requires that the seller legally
owns the goods. These provisions are modeled by the
preconditions function $Pre$. It specifies for each action $a$ the
conditions in the underlying static models that need to obtain for $a$
to be executable in the first place.

The key to our modeling of deontic or legal dynamics is the relation
$\leq_{j\to k}^{\A_i}$ and the post-condition function $Post$. The first
encodes $i$'s potential for changing the ideality ordering concerning $j$'s obligations towards $k$, in states satisfying certain preconditions. It thus captures, so to
speak, the deontic potential or deontic effectivity of $i$'s actions in
$A$ on the different directed ideality relation. The second captures the legal ability to render certain legal facts true~\citep{herzig2011dynamic}. For instance, in our example, issuing a parking ticket thereby
makes it obligatory that Ivy pays a fine. The situation here involves both deontic changes on the legal potential as well as on the truth of legal facts.

Note that if we were to adopt the same strategy as in dynamic
epistemic logic to interpret the relations $\leq_{j \to k}^{\A_i}$, we would
read $a_1 \leq_{j \to k}^{\A_i} a_2$ in terms of comparative deontic or legal
ideality. That is we would say that $a_2$ is at least as ideal,
legally speaking, as $a_1$, as far as $j$'s obligations towards $k$ are concerned. In the epistemic interpretation the
ordering between events is indeed interpreted in the same way as the
ordering between states in the static models, that is in terms of
comparative plausibility~\citep{baltag2008qualitative}. If, in that
context, we have for instance $e_1 > e_2$, it means that $e_1$ is
strictly more plausible than $e_2$. 

This reading, however, can be misleading in the present deontic
interpretation. Saying that a deontic action $a_2$ is at least as, or
even more ideal than another action $a_1$, legally speaking, strongly
suggests that performing $a_2$ is at least as good as performing
$a_1$. If we accept the principle that everything at least as ideal as
a permissible action is also permissible, we would then arrive at $a_2$
becoming legally permissible as soon as $a_1$ is. As we will see in
Section~\ref{sec:app1} this is not the case when legal ability and
legal permissibility come apart. In such cases some agents may be in
a position to perform two actions, say $a_1$ and $a_2$, the effects of
which would be to make all states that satisfy the precondition of
$a_2$ strictly higher in the relation $\leq_{j \to k}$ than all states that satisfy the precondition of
$a_1$. This is naturally encoded by putting $a_1 <_{j\to k}^{\A_i} a_2$. But
when this legal ability does not entail legal permissibility, it could
be that $a_1$ is permissible while $a_2$ is not. In such cases it is
not clear to what extent one can read $a_1 <_{j \to k}^{\A_i} a_2$ as ``$a_2$ is
legally more ideal'' or ``better than $a_1$''. 

For this reason we use the more neutral reading of $\leq_{j\to k}^{\A_i}$ in
terms of deontic potential or legal effectivity. In most cases, of
course, legal ability coincides with legal permissibility, and when
this happens one could add the additional evaluative layer to the
interpretation of the relation $\leq_{j\to k}^{\A_i}$. The reader should keep
in mind, however, that this interpretation only works in those
cases. 

In order to improve readability we index deontic-actions models with particular agents. This is simply a mnemonic device, and plays no role in the mathematical definitions below. Conceptually, however, this allows us to distinguish between different legal competences of different agents, \emph{over the same static model}, i.e. in a given legal situation. Indeed,  some agents, for instance the judiciary, might have, in a given situations, competences that are different than those of a layperson in that same situation. These could in principle affect the legal relation of the same individuals, only in different ways. A layperson might have the competence to enter or renew a contract between her and someone else, but only the judiciary might have the competence to rule a contract as invalid. In that case we would model this using two different action models, one for the judiciary and one for the layperson, and index them accordingly. In Section~\ref{sec:2.4} we present an example with two different action models for two different agents, which can be executed in the same circumstances. Once again, however, the reader should keep in mind that these particular indices on action models do not play any role in the definitions below.

Observe, furthermore, that the relation $\leq_{j\to k}^{\A_i}$ is indexed by a pair of agents $j, k$ which might differ from $i$. The reason for this is that it is possible for some agents to change the legal relations between two different parties. A common example is civil servants with the legal power to marry couples. A specific deontic action model $\mathcal{A}_i$ for such an agent $i$ thus captures not only her potential to change the legal relations where she is the subject or the addressee, i.e. where either $i \to j$ or $j \to i$ for some $j$, but also the potential of $i$ to change the legal relations of other pairs $j,k$ of agents different from her. This modeling choice is in line with a number of recent proposals to capture  legal relations~\citep{lorini2008logical,herzig2011dynamic}, but notably generalizes~\citep{van2016modelling}, where the legal competences of an agent $i$ are restricted to the relation that $i$ has towards others. 

\subsection{Deontic Action Models: an example}\label{sec:4.2-1-2}
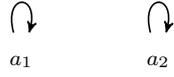
\begin{figure}[hbt!]\centering
	
	\begin{tikzpicture}[->,>=stealth',shorten >=1pt,shorten <=1pt,
	auto,node distance=1cm,semithick,scale=.6]
	\tikzstyle{every state}=[draw=none,text=black]
	\node[state,label=left:{}] at (1,1) (a1)
	{{\small $\textcolor{black}{a_1}$}};
	\node[state,label=right:{}] at (4,1) (a2) {{\small
			$a_2$}};
	
	\path (a1) edge [loop above] node {} (a1);
	\path (a2) edge [loop above] node {} (a2);
	\end{tikzpicture}
	\caption{The deontic action model $\A_{John}$ for John the parking officer. The pre- and post-conditions of $a_1$ and $a_2$ are given as
		follows: $Pre(a_1): \neg d \wedge p, Post(a_1)(f) = \top $,
		$Pre(a_2): d \vee \neg p, Post(a_2)(f) = \bot $. There are only
		reflexive $\leq^{\A_{John}}_{i \to c}$ arrows in this
		deontic action model. }\label{fig:4.2}
\end{figure}

Let us now turn to an example to illustrate this. John is a parking officer. He can confirm an illegal parking by issuing a parking ticket, if indeed a car is parked without a permit. To model this we use two deontic actions, $a_1$ and $a_2$. The first encodes confirming that a car is parked without a permit displayed. This is expressed, on the one hand, in the
precondition $\neg d \wedge p$ and, on the other hand, in the
post-condition that after $a_1$ is executed $f$ becomes true. The
second action $a_2$ captures the case where he does not confirm a
violation of the parking regulations. This deontic-action model is illustrated in
Figure~\ref{fig:4.2}. 


\subsection{Updating with Deontic Action Models}
The effect of executing a deontic action in a particular state is
computed by the so-called lexicographic update. The intuitive idea is
the following. First, a particular deontic action can only be executed
in states where its preconditions hold. Second, the normative or legal
status of a particular state after the update is a function of the
normative status of that state before the update, together with the deontic
potential or legal effectivity of the actions that can be executed in that
state. The update gives priority to the effect of the deontic
actions, hence the label ``lexicographic.'' This reflects the idea
that successfully enacting a legal competence means effectively
changing the underlying legal relations. Third, and finally, executing
a deontic action only changes legal relation and legal fact. The
non-deontic actions remain unchanged after the update. Formally, this
gives the following:

\begin{definition}\label{lexico-model} Let $\M$ be a preference-action model and $\A_i$ be a deontic action model. The preference-action model 
	$\M \otimes\A_i = \langle W', \{\leq'_{j\to k}, \sim'_j\}_{j,k \in \Agt}, V'
	\rangle $ is defined as follows, for all agents $j,k$.
	\begin{itemize}
		\item $W' = \{(w, a) \mid \M, w \models Pre(a) \text{, where }a \in A\}$.
		\item $(w, a) \leq_{j \to k}' (w', a')$ iff either $a <_{j \to k}^{\A_i} a'$ or $a \cong_{j \to k}^{\A_i} a'$ and $w \leq_{j \to k} w'$.
		\item $(w,a) \sim'_j (w', a')$ iff $w \sim_j w'$ 
		\item $ V'(p) = \{(w,a) \in W': \M, w \models Post(a)(p) \}$.
	\end{itemize} 
\end{definition}

The lexicographic update thus takes pairs of preference-action models
and deontic action models as input and returns an updated model
$\M \otimes \A_i$.  The domain of this new model is the set of pairs
$(w, a)$ such that $\M, w$ satisfies the precondition of $a$. This
captures the idea that actions are only executed in states where their
preconditions hold. As mentioned, the adjective ``lexicographic'' comes
from the update rule for the preference order $\leq_{j \to k}'$. It
gives priority to the legal effectivity of deontic
actions.\footnote{Our definition of the update rule is slightly
	different from the standard one, used for instance
	in the proposals of~\cite{baltag2008qualitative,benthem2014priority}. Here two pairs
	$(w, a)$ and $(w', a')$ are connected in the updated model as soon
	as $a >^{\A_i}_{j \to k} a'$ or the other way around. This is so
	irrespective of whether $w$ and $w'$ were initially connected by
	$\leq_{j \to k}$. We made this modeling choice because it allowed us to capture more naturally some of the examples that we present here.\label{ftn:update} }

Lexicographic updates capture changes at the legal or deontic level,
taken in isolation. Borrowing from the epistemic interpretation of
this formalism, we could call them ``pure'' deontic
actions~\citep{van2007dynamic}. This is encoded in the condition
defining the valuation $V'$ and the equivalence relation in the
updated model:
\begin{align*}
(w, a) \in V'(p) &\Leftrightarrow w \in V(Post(a)(p)),\\
(w,a) \sim_j' (w', a')  &\Leftrightarrow w \sim_j w'.
\end{align*}
In other words, executing a deontic action, or enacting a legal competence, first does not change the non-deontic powers of the agents. This is what the second line above specifies. Second, it changes the facts only to the extent specified by the deontic actions themselves. This is what the first line encodes. One can take such pure deontic actions to be actions that are explicitly
defined by the legislator, for instance entering into a contract or
getting married. These are the ``purely legal'' actions, which Hohfeld
himself took care to distinguish from other types of changes:
\begin{quote}
	At the very outset it seems necessary to emphasize the importance of
	differentiating purely legal relations from the physical and mental
	facts that call such relations into being.~\citep[p.20]{hohfeld1913some}
\end{quote}
Pure deontic actions will, of course, usually be generated by non-deontic
ones, in Goldman's sense of ``generation''~\citep{goldman1970theory}. Changes brought about by deontic
actions will usually supervene on non-deontic changes. Entering a
contract might require signing certain documents, or getting married
uttering some words. The models proposed by~\cite{jones1996formal,governatori2005temporalised,grossi2006classificatory} capture such conventional or legal
action generation. The focus here is on the changes at the level of
the generated deontic actions, in isolation from their non-deontic
counterparts. A full comparison and combination between our model of
deontic actions and those other models is left for future work.

\subsection{Updating with Deontic Action Models: An Example}\label{sec:2.4}
\begin{figure}[htbp]
	\centering
	\begin{tikzpicture}[->,>=stealth',shorten >=.5pt,shorten <=.8pt,
	auto,node distance=.8cm,semithick, scale=.8]
	\tikzstyle{every state}=[draw=none,text=black]
	
	
	\node[state,label=left:{$\neg d, p, f$}] at (0,0) (1) {{\small $\textcolor{black}{(w_1, a_1)}$}};
	\node[state,label=left:{$d, p, \neg f$}] at (0,4) (2) {{\small $(w_2, a_2)$}}; 
	\node[state,label=right:{$d,\neg p, \neg f$}] at (4,4) (3) {{\small $(w_3, a_2)$}};
	\node[state,label=right:{$\neg d, \neg p, \neg f$}] at (4, 0) (4) {{\small $(w_4,a_2)$}};
	
	\path[draw,dashed,thick] (-.8,3.5) rectangle (0.9,4.5);
	
	\path[draw,dashed,thick] (-.8,-.5) rectangle (0.9,0.5);
	
	\path[draw,dashed,thick] (3.1,3.5) rectangle (4.8,4.5);
	
	\path[draw,dashed,thick] (3.1,-.5) rectangle (4.8,.5);
	
	\path (3) edge[<->, draw=black] (4); 
	\path (2) edge[<->, draw=black] (3);
	\path (1) edge [loop above] node {} (1);

	\end{tikzpicture}
	
	\caption{The model $\M \otimes\A_{John} $ resulting from John's
		execution of his deontic actions. For clarity the reflexive
		arrow at $(w_1,a_1)$ is displayed, but all the others are
		omitted. }\label{fig:4.3}
\end{figure}
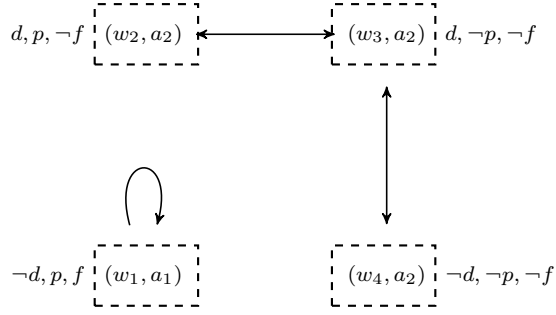

With this in hand we can return to our running example.  The legal
effects of enacting John's legal competence in Ivy's situation are
represented by the updated model in Figure~\ref{fig:4.3}. John's
confirmation of the fact that Ivy has parked illegally leads to the removal of all arrows from the
state where she is parked without a permit, i.e. $(w_1, a_1)$ to all
others. That action also changes the truth value of the legal facts that Ivy pays the fine in the now most ideal worlds, legally speaking. In $(w_1, a_1)$ the city has now an unconditional claim
against her to pay a fine.\footnote{The semantics of conditional obligation that we use also yields, after the update, that at $(w_1, a_1)$ the city has an unconditional claim against Ivy to park and not display a permit. This counter-intuitive prediction is a consequence of the fact that this semantics validates $O_{i \to c}(\phi/\phi)$, for any $\phi$. This means that, when the current state can only see itself according to the relation $\leq^{\A_{John}}_{i \to c}$, we obtain that  $O_{i \to c}\phi$ whenever $\phi$ is true. This is an issue that is shared by most preferential accounts of conditional obligation~\citep{zvolenszky2002possible}, and that also affects some approaches based on default logic~\citep{fuhrmann2017deontic}. Addressing this thoroughly would require using a completely different approach to static conditional obligations, which would go beyond the scope of this paper. Note, however, that this is a defect that only affects the static conditional obligations. One could also interpret "given $\phi$, $\psi$" dynamically, for instance using in terms of public announcements of the form $[\phi!]\psi$, which are a special cases of the updates that we define here~\citep{baltag2008qualitative}. It is well known that $[\phi!]\phi$ is \emph{not} a valid formula~\citep{van2007dynamic}.}  In all other states, i.e. where Ivy either does not park or else displays the permit, her duties and privileges, both conditional and unconditional, remain unchanged. 

\begin{figure}[hbt!]\centering
	
	\begin{tikzpicture}[->,>=stealth',shorten >=1pt,shorten <=1pt,
	auto,node distance=1cm,semithick,scale=.6]
	\tikzstyle{every state}=[draw=none,text=black]
	\node[state,label=left:{}] at (1,1) (b1)
	{{\small $\textcolor{black}{b_1}$}};
	\node[state,label=right:{}] at (4,1) (b2) {{\small
			$b_2$}};
	\path (b1) edge[->, draw=black] (b2); 
	\end{tikzpicture}
	\caption{The deontic action model $\A_{Mary}$ for Mary, John's
		superior at the town hall. The pre- and post-conditions of
		$b_1$ and $b_2$ are given as follows:
		$Pre(b_1): \neg d \wedge p, Post(b_1)(f) = \bot $,
		$Pre(b_2): d \vee \neg p, Post(b_2)(f) = \bot $. There are only
		reflexive $\leq^{\A_{Mary}}_{i \to c}$ arrows in this
		deontic action model, all the others being omitted. }\label{fig:mary}
\end{figure}
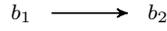

This analysis of John's legal power is close to the one provided
by~\cite{markovichSL} in terms of public announcement, with one
important difference. Here John's actions are ``soft'' in the sense
that they are reversible. Suppose for instance that Mary, John's
superior at the town hall, has the power to cancel John's
decision, i.e. to nullify his observation, and by that very fact
restore Ivy's original situation, which in particular cancels the
city's claim against her to pay a fine. This can be represented by
the deontic action model in Figure~\ref{fig:mary}. It is
straightforward to check that updating $\M \otimes \A_{John}$ with
$\A_{Mary}$ yields a model which is isomorphic to $\M$. This kind of
reversibility is ruled out by public announcements since they
completely remove states. 

To express the effect of deontic action the language $\mathcal{L}$ is
extended with a dynamic, unary operator $[\A_i, a]$, with the following semantics:
\begin{itemize}
	\item $\M, w \models [\A_i, a] \phi$ iff if $\M, w \models Pre(a)$
	then $\M \otimes \A_i, (w, a) \models \phi$.
\end{itemize}
A formula $[\A_i, a] \phi$ thus is read as ``$\phi$ holds after
executing $a$'' or, more precisely, ``whenever $a$ is executable,
i.e. its preconditions hold, $\phi$ holds after executing $a$''. These
dynamic modalities have duals, which we write
$\langle \A_i, a \rangle$. A formula of the form
$\langle \A_i, a \rangle \phi$ should be read as ``$a$ is executable
and $\phi$ holds after one of its executions.'' So the only difference between the dynamic ``box" $[\A_i, a]$ and its ``diamond" $\langle \A_i, a \rangle$ is that the latter states that the preconditions of $a$ hold in the current state, while the former only describes what happens \emph{if} they hold. In cases where the preconditions are false we obtain $[\A_i, a] \phi$ true for every $\phi$, while every such $\langle \A_i, a \rangle \phi$ turns out \emph{false}. In particular, $\langle \A_i, a \rangle \top$ can be read as saying that $a$ is executable in the current state, while $[\A_i, a ] \bot$ that $a$ is not executable. Also, importantly, reading $\langle \A_i, a \rangle \phi$ as ``some executions of $a$ result in $\phi$" might be misleading. This is so because the result of updating any preference-action model with a deontic action model is unique. In other words, updates are deterministic. 

\subsection{Power and Immunity}

We define power and immunity using these dynamic modalities. To do
this we use Lindhal's notation~\citep{lindahl1977position} and write
$T(i,j,\psi/\phi)$ for an arbitrary normative position, static or dynamic.
\begin{definition}[Power and Immunity: Global Version]
	Let $T(j,k,\psi/\phi)$ be a normative position, $\M$ a preference-action model and $w$ any state in $W$. Then:
	\begin{itemize}
		\item $i$ has a \emph{power} against $j,k$ regarding
		$T(j,k,\psi/\phi)$ at $\M, w$:
		$$\M, w \models \bigvee_{a \in A}\langle \A_i,a\rangle T(j,k,\psi/\phi)$$
		\item $j,k$ have an \emph{immunity} against $i$ regarding
		$T(j,k,\psi/\phi)$ at $\M, w$:
		$$\M, w \models \neg \bigvee_{a \in A}\langle\A_i,a\rangle T(j,k,\psi/\phi)$$
	\end{itemize}
\end{definition}
Before we discuss the local/global distinction, three comments
regarding this definition are in order. First, we define legal
competences as tripartite relations. We say that $i$ has a power
against two agents $j$ and $k$ regarding the normative position
$T(j,k,\psi/\phi)$ whenever there is an \emph{executable} -- hence the
use of the modality $\langle \A_i, a \rangle$ -- deontic action of $i$
which would result in $T(j,k,\psi/\phi)$. On the other hand we define
the fact that $j, k$ have an immunity against $i$ regarding
$T(j,k,\psi/\phi)$ if $i$ does not have \emph{any} action that
would result in the position $T(j,k,\psi/\phi)$. The reason for this
tripartite definition is, again, that we want to be able to capture cases where a particular agent, for instance the
judiciary or an institutional representative, might have the competence
to change the legal relations between other agents, for instance by
declaring someone guilty and thereby creating a claim to
reparation. This is indeed the case in our running example above. Of
course, this definition allows for cases where, for instance, $i$ has
the legal power to create a claim right herself against $k$. In
that case we simply have $i = j$ in the above definition.

Second, this definition captures power as \emph{potential} to change legal relations, and thus does not conflate this potential with the actual exercising of the power. To be sure, our definition ensures the precondition necessary for a power to hold. We are indeed using the ``diamond" $\langle \A_i,a\rangle$, instead of the ``box" $[\A_i,a]$ to define power. This does not mean, however, that the power is actually exercised in that state. The formula only describes what would be the potential consequences of exercising it.  

Third, this definition captures the Hohfeldian square of legal
competences (Table~\ref{legrig}, page~\pageref{legrig}). The
correlative of agent $i$'s power against $j,k$ regarding $T(j,k,\phi)$ 
is $j$ and $k$'s liability against $i$ regarding the same static
normative position. In classical formalizations of Hohfeldian rights,
for instance~\citep{kanger1966rights}, lateral moves in the diagrams of
Table~\ref{legrig} are modeled as changes in the directionality of the
corresponding rights. Liability is modeled in the same way as its correlative, $i$'s
power: $$j,k \text{ have a liability against }i\text{ regarding } T(j,k,\phi) \text{ iff } \bigvee_{a \in A}\langle \A_i,a\rangle T(j,k,\phi)$$
Vertical moves in the same diagrams, i.e. moving to opposites, are
modeled by changing the polarity of the corresponding right. So the
opposite of $j,k$'s liability corresponds to
$$\neg \bigvee_{a \in A}\langle \A_i,a\rangle
T(j,k,\phi)$$ This is exactly our definition of $j$ and $k$'s
immunity against $i$ regarding $T(j,k,\phi)$. Pushing the
negation inside we obtain:
$$\bigwedge_{a \in A} [\A_i,a] \neg T(j,k,\phi)$$ 
This states that no deontic action of $i$ would result in
$T(j,k,\phi)$. Again, since directionality is left implicit, the
same formula corresponds to moving left in the diagram of
page~\pageref{legrig}, and gives a natural formalization of $i$'s
no-power against $j,k$ regarding $T(j,k,\phi)$.

This definition of power and immunity has, however, an important
shortcoming. It predicts that $i$ has a power against $j,k$ regarding
$T(j,k,\psi/\phi)$ as soon as $i$ has an executable deontic action
that \emph{preserves} or, in other words, does not change
$T(j,k,\psi/\phi)$. This can be seen in our running example by looking
at what happens in states other than $w_1$. There Ivy's conditional
obligation to display her permit whenever she parks remains unchanged
by John issuing a ticket. $O(Do_{Ivy}d/p)$ holds in all states except
$w_1$ before and after the update. So the global definition of power
yields that John has a power against Ivy regarding her
obligation to display a ticket, which might be seen as counter-intuitive, since John's actions do not change that right. We
furthermore obtain the reverse predictions for immunity. Cases where none of
$i$'s actions result in $T(j,k,\psi/\phi)$ are compatible with
some of them effectively changing $T(j,k,\psi/\phi)$ to
$\neg T(j,k,\psi/\phi)$. This, again, appears rather
counter-intuitive.

A natural way to make up for this shortcoming is to enforce that
power induces genuine changes in the underlying normative positions, while
immunity genuinely preserves them. To do this we move to a
\emph{local} definition of legal competences, in the sense that it is
relative to a given preference-action model $\mathcal{M}$ and state
$w$ in it.
\begin{definition}[Power and Immunity: Local Version] Let
	$\mathcal{M}$ be a preference-action model and $w$ a state in it such
	that $\mathcal{M}, w \models T(j,k,\psi/\phi)$.
	\begin{itemize}
		\item $i$ has a \emph{power} against $j, k$ regarding
		$T(j,k,\psi/\phi)$ at $\mathcal{M}, w$ iff 
		$$\mathcal{M},w \models \bigvee_{a \in A} \langle \A_i,a \rangle \neg T(j,k,\psi/\phi)$$
		
		\item $j, k$ have an \emph{immunity} against $i$ regarding
		$T(j,k,\psi/\phi)$ at $\mathcal{M}, w$ iff: 
		$$\mathcal{M}, w \models \bigwedge_{a \in A} [\A_i,a]T(j,k,\psi/\phi)$$
		
	\end{itemize}
\end{definition}
This definition is local in the sense that it is relative to a given
pointed preference-action model. It requires checking whether the
given right $T(j,k,\psi/\phi)$ holds at $w$ in the underlying
preference-action model. For $i$ to have a power regarding that right
we require that $i$ has some executable deontic action that, when
executed, would change the polarity of that right. So $i$ has a power to establish a
position $T(j,k,\psi/\phi)$ if it doesn't hold in the given situation, but
$i$ can make it true. Similarly, $i$ has a power to cancel a position if
that position holds in the given situation, though $i$ can make it
false.\footnote{This view of legal powers as potential to \emph{change} or "flip", so to speak, the truth value of certain legal facts is also used by~\citet{van2016modelling}. Note, however, that in that paper they restrict to the case where the agents have control over atomic propositions, while where the legal powers extend to any right  $T(j,k,\psi/\phi)$.} For immunity we require the contrary, namely that no executable
action changes the static normative position in question. Both
notions are context dependent. Whether they hold depends
on the specifics of the underlying static legal relations, hence the
local definition, relative to a given model-state pair.

This alternative definition is again tripartite and keeps the
correlative and opposite relation of the Hohfeldian diagram. To see
this latter point, observe that negating the fact that $i$ has a power
against $j, k$ regarding $T(j,k,\psi/\phi)$ at $\mathcal{M}, w$ in a situation where $T(j,k,\psi/\phi)$ does hold gives
us:
$$\mathcal{M},w \models \bigwedge_{a \in A}[ \A_i,a] T(j,k,\psi/\phi)$$
which is precisely the definition of immunity. This states that no
executable action of $i$ will change the status of
$T(j,k,\psi/\phi)$, which gives the ``no-power'' correlative of immunity. 

This new definition avoids the shortcoming of the global
definition. On the one hand it rules out cases of trivial powers that
do not change the underlying legal relations. In our running example it
yields the intuitively correct prediction that John has no power
against Ivy regarding the unconditional obligation to display her
parking permit or, equivalently, that Ivy has an immunity against John
regarding that obligation. 

The local definition is of course context- or model-dependent, in the sense that it is relative to a concrete legal situation, as opposed to constituting a general definition of what ``having a legal power towards  $T(j,k,\psi/\phi)$'' means. It provides an analysis of \emph{tokens} of legal abilities, as opposed to types.   

As such, the local definition comes with stronger logical relations
between different powers than the global one. The latter is, for
instance, compatible with an agent $i$ having both a power regarding
$T(j,k,\psi/\phi)$ and a power regarding its negation. This combination is
inconsistent under the local definition. These two local powers require that we
consider situations where $T(i,j,\psi/\phi)$ is, respectively, false and
true, which of course cannot occur simultaneously. This highlights
the fact that this local definition of power boils down to modeling
power as the ability to \emph{establish} or \emph{cancel} a certain
legal relation. The former assumes that the legal relation does not
hold before the power is enacted, and the latter that it does. 

A welcome corollary of this is that abstaining from enacting a power
is distinct, in this logic, from having a power to the
contrary. As we have just argued, a locally defined power regarding $T(i,j,\psi/\phi)$
is incompatible, in this model, with a power regarding
$\neg T(i,j,\psi/\phi)$. This is so because the former models a case of
establishing $T(i,j,\psi/\phi)$ and the latter of canceling $T(i,j,\psi/\phi)$,
and the two have inconsistent preconditions. Either of these
competences, however, are compatible with the ability to refrain from
exercising them, i.e. to do nothing at all. This ability can be modeled
by adding to any action model an action unconnected by $\leq^{\A_{i}}_{j \rightarrow k}$
to all others except itself, and with $\top$ as precondition. Updating
with that larger deontic action model will create a copy of the
original model alongside the updated one, and of course in that model
the original legal relations remain unchanged.

This formalization of dynamic rights has two assets in comparison with
classical, reductive approaches. First, it explicitly captures, both
semantically and syntactically, the dynamic character of power and
immunity. Second, as we will see below, this clear static-dynamic
distinction allows for a natural distinction between legal ability and
legal permissibility.

\subsection{Axiomatization} \label{sec:system} 

Axiomatizing the set of validities for the frames and updates
just defined proceeds in two modular steps. First the validities for the static modalities of $\mathcal{L}$ are axiomatized, and second provide an axiomatization of  the dynamic extension. For the static part
the axiomatization proceeds in a standard manner. We use {\bf S4}
for $[\leq_{i \to j}]$ and {\bf S5} both for $U$ and $Do_i$. Interaction between
$[\leq_{i \to j}]$, $Do_i$ and $U$ can be captured by standard inclusion
axioms. Of course, each modality satisfies the necessitation rule. See
Table~\ref{tb:axiomstatic}.

\begin{table}[htbp]
\centering
	\fbox{
		\parbox{.97\textwidth}
		{ S4 for $\leq$: \\
			\begin{tabular}
				{p{0cm}p{10cm}}
				&$\vdash $ $ [\leq_{i \to j}](\phi \rightarrow \psi) \rightarrow [\leq_{i \to j}] \phi \rightarrow [\leq_{i \to j}] \psi$ \\ 
				&$\vdash $ $ [\leq_{i \to j}] \phi \rightarrow \phi$ \\
				&$\vdash $ $ [\leq_{i \to j}] \phi \rightarrow [\leq_{i \to j}][\leq_{i \to j}]\phi$ \\ 
				& From $\vdash \phi $ infer $ \vdash [\leq_{i \to j}]\phi$ \\
			\end{tabular}\\\;\\
			S5 for $U$, and similarly for $Do_i$:  \\
			\begin{tabular}
				{p{0cm}p{10cm}}
				&$\vdash $ $ U(\phi \rightarrow \psi) \rightarrow U \phi \rightarrow U \psi$ \\ 
				&$\vdash $ $ U \phi \rightarrow \phi$ \\
				&$\vdash $ $ U \phi \rightarrow U U \phi$ \\
				&$\vdash $ $ \neg U \phi \rightarrow  U \neg U \neg \phi$ \\ 
				& From $\vdash \phi $ infer $ \vdash U\phi$ \\
			\end{tabular}
			\\\;\\
			Interaction axioms:  \\
			\begin{tabular}
				{p{0cm}p{10cm}}
				&$\vdash $ $ U \phi \rightarrow [\leq_{i \to j}] \phi$ \\
				&$\vdash $ $ U \phi \rightarrow Do_i \phi$ \\
			\end{tabular} 
			\caption{\small Axioms for preference-action models\label{tb:axiomstatic}} 
	}}
	\end{table} 

Axiomatizing the dynamic part uses the well-known ``reduction axioms''
methodology~\citep{baltag2008qualitative,benthem2014priority}. Formulas
containing dynamic modalities are shown to be semantically equivalent
to formulas of $\mathcal{L}$, that is without dynamic modalities.  The
formulas in Table~\ref{tb:axiom} indeed show how to ``push'' dynamic
modalities inside the various connectives and modal operators of the
static language, until they range over atomic propositions where they
can be eliminated.\footnote{These axioms are by and large standard,
	c.f. again~\citep{baltag2008qualitative,benthem2014priority}. The
	only difference comes from our slightly non-standard clause in the
	lexicographic update rule, which results in the use of the universal
	modality in the first group of conjuncts in the second equivalence
	from below. See again footnote~\ref{ftn:update} on
	page~\pageref{ftn:update} for details.}
\begin{table}[hbt!]
	\fbox{
		\parbox{.97\textwidth}
		{
			\begin{tabular}
				{p{0cm}p{10cm}}
				&$\vdash $ $ [\A_i, a] p \leftrightarrow (Pre(a) \rightarrow Post(a)(p))$ \\
				&$\vdash $ $ [\A_i, a]  \neg \phi \leftrightarrow (Pre(a) \rightarrow \neg[\A_i, a] \phi)$ \\
				&$\vdash $ $[\A_i, a] (\phi \wedge \psi) \leftrightarrow [\A_i, a]\phi \wedge [\A_i, a] \psi $\\
				&$\vdash $ $ [\A_i, a]U\phi \leftrightarrow (Pre(a) \rightarrow U[\A_i, a]\phi) $\\
				&$\vdash $ $ [\A_i, a] [\leq_{j \to k}] \phi \leftrightarrow (Pre(a) \rightarrow $\\
				& \quad\quad\quad\quad\quad\quad\quad\quad\quad\quad\quad $\bigwedge_{c >^{\A_i}_{j \to k} a} U[\A_i, c]\phi \wedge \bigwedge_{c\cong^{\A_i}_{j \to k} a}[\leq_{j \to k}][\A_i, c]\phi) $\\
				& $\vdash $ $ [\A_i, a] Do_j \phi \leftrightarrow  (Pre(a) \rightarrow  Do_j [\A_i, a]\phi) $\\
			\end{tabular} 
			\caption{\small Reduction Axioms for Lexicographic Update} \label{tb:axiom}
	}}
\end{table} 

These formulas are sound with respect to the lexicographic update over preference-action models. Taking them as axioms thus makes formulas containing dynamic modalities provably equivalent to formulas of $\mathcal{L}$. Completeness for the extended language then follows
from completeness of the static part with respect to the class of
preference-action models. This is a standard technique to prove completeness for dynamic extensions of static languages. See e.g.~\cite[p.196]{van2007dynamic} for details.
\begin{theorem}
	The axioms in Table~\ref{tb:axiomstatic} together with the
	reduction axioms in Table~\ref{tb:axiom}, all propositional
	tautologies and Modus Ponens are sound and complete with respect to
	the class of preference-action frames and lexicographic updates. 
\end{theorem}

\subsection{Reduction of Legal Competences to Static Legal Relations} \label{sec:reduction} 

Through the soundness of the reduction axioms, together with the fact
that conditional obligations are definable in $\mathcal{L}$, we obtain
that power and immunity are also reducible to the language where static normative
positions are defined. So the approach presented here is reductive. This
reduction, however, is complex, especially in comparison with the reduction proposed
for instance by~\cite{lindahl1977position}. 
\begin{multline*}
 [\A_i, a]O_{j\to k}(\psi / \phi) \leftrightarrow \\
\bigwedge_{b \in \max(|\unll a,\varphi|)}U O(\bigwedge_{d \in \uncong[b]}(\langle\A_i, d \rangle \varphi \rightarrow \langle\A_i, d \rangle \psi)/\bigvee_{c \in \uncong[b] } \langle\A_i, c \rangle \varphi) \wedge\\
 \bigwedge_{b \in |\uncong a, \varphi| }O(\bigwedge_{d \in \uncong[b]}(\langle\A_i, d \rangle \varphi \rightarrow \langle\A_i, d \rangle \psi)/\bigvee_{c \in \uncong[a]}\langle\A_i, c \rangle \varphi) 
\end{multline*}
where $|\unll a,\varphi| =\{ c >^{\A_i} a   \mid Post(c)(\varphi) = \top \}$ and $|\uncong a,\varphi| =\{ c \cong^{\A_i} a   \mid Post(c)(\varphi) = \top \}$.

The
complexity of this formula results from it essentially encoding
syntactically the lexicographic update rule in combination with the
specific semantic definition of obligations as truth in all the
most ideal worlds. For unconditional obligations we obtain the following, slightly more readable formula.

\begin{multline*}
 [\A_i, a]O_{j\to k}\psi  \leftrightarrow \\
\bigwedge_{b \in \max(|\unll a,\varphi|)}U O(\bigwedge_{d \in \uncong[b]}(Pre(d) \rightarrow \langle\A_i, d \rangle \psi)/\bigvee_{c \in \uncong[b] } Pre(c)) \wedge\\
 \bigwedge_{b \in |\uncong a, \varphi| }O(\bigwedge_{d \in \uncong[b]}(Pre(d) \rightarrow \langle\A_i, d \rangle \psi)/\bigvee_{c \in \uncong[a]}Pre(c)) 
\end{multline*}

Each formula show that the effects of changes in legal relations are
reducible to statements describing legal relations holding
\emph{before} the deontic action takes place.

In view of the complexity of the resulting formulas we cannot, however, claim that the present approach yields much insight into the relation between static and dynamic rights. Unlike non-reductive approaches, this deontic re-interpretation of dynamic-epistemic logic \emph{does} come with a rich theory combining the static and the dynamics parts. Whether this richness can be cashed out into concrete insights for the Hohfeldian typology remains however to be seen.

\section{Legal Ability and Legal Permissibility}
\label{sec:app1}
Although legal ability and legal permissibility often go together,
they are conceptually distinct
notions~~\citep{makinson1986formal,jones1996formal}. The German Civil
Code ({\it B\"{u}rgerliches Gesetzbuch}) illustrates this through ``Missbrauch der Vertretungsmacht'', which we translate freely here as ``abuse of the terms of representation''. This is defined as follows:
\begin{quote}
    A failure to comply with restrictions from the legal relationship on which the power of representation is based when undertaking a legal transaction on behalf of the representative. Since the power of representation is abstracted from the underlying legal relationship, \emph{such restrictions do not lead to a restriction of the power of representation}, so that \emph{even a declaration made in abuse of the power of representation remains within the scope of the power of representation}. \footnote{Our emphasis. Freely tranlated from ``Nichteinhaltung von Beschränkungen aus dem der Vertretungsmacht zugrunde liegenden Rechtsverhältnis bei Vornahme eines Rechtsgeschäfts in Stellvertretung durch den Vertreter. Da die Vertretungsmacht von dem zugrunde liegenden Rechtsverhältnis abstrakt ist, führen solche Beschränkungen nicht zu einer Einschränkung der Vertretungsmacht, so dass sich auch eine unter Missbrauch der Vertretungsmacht abgegebene Erklärung grundsätzlich im Rahmen der Vertretungsmacht hält.'' Source \url{http://rechtslexikon.net/d/missbrauch-der-vertretungsmacht/missbrauch-der-vertretungsmacht.htm}}
\end{quote}

The idea being that provision appears to protect the counter-party of a transaction that violates the terms of a representation contract with a principal.\footnote{The German Civil Code of course explicitly voids this provision if there was proven collusion between buyer and seller, or when the seller should have known the terms of the agent's contract.} Let us illustrate this by a simple example. 

Suppose that $i$ contracts $j$ to buy supplies for her house. Our agent $i$ adds the condition not to spend more than 500 euro, if something is bought at all. 
Agent $j$ goes to the shop $k$ and buys goods for 600 eurso. The sale is
a deontic action made by $j$ in $i$'s name and is
valid although it violates the ``legal relationship'' between $i$ and $j$, i.e. the contract. After the sale $i$ owns the good and owes 600 euro to $k$. Of course, the German Civil Code might have additional provisions regulating reparations in case of breach of contract. We bracket them here, as our goal is to illustrate the difference between legal ability and legal permissibility. So $j$'s buying from
$k$ is not legally permissible, because excluded by the contract between $i$ and $j$, while being legally feasible for $j$. We now show that our logical model of legal competences can capture this case in a simple manner.

\subsection{Agent $j$'s Power to Buy} \label{sec:legab}
Ownership rights involve a number of different obligations and
permissions, but for simplicity here we represent them as the fact
that the owner of a given good is in possession of it. So let $p$ be the fact that $i$ is in possession of the goods for sale at $k$, and $\neg p$ the fact that $k$ is in
possession of those goods. Let furthermore $f$ be the
fact that $i$ pays the 600 euro to $k$.

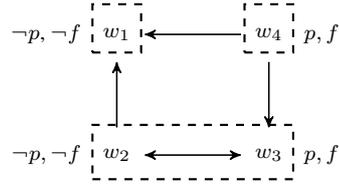
\begin{figure}[htbp]
	{ \centering
		\begin{tikzpicture}[->,>=stealth',shorten >=.1pt,shorten <=.1pt,
		auto,node distance=.2cm,semithick, scale=.8]
		\tikzstyle{every state}=[draw=none,text=black]

		\node[state,label=left:{$\neg p$, $\neg f$}] at (0,4) (1)
		{{\small $w_1$}};
		\node[state,label=left:{$\neg p, \neg f$}] at (0,2) (2)
		{{\small $w_2$}};
		
		\node[state,label=right:{$p, f$}] at (2.5,4)
		(4) {{\small $w_4$}};
		\node[state,label=right:{$p, f$}] at (2.5,2)
		(3) {{\small $w_3$}};

		\path (1) edge[<-, draw=black]  (2);
		\path (4) edge[->, draw=black]  (3);
		\path (3) edge[<->, draw=black]  (2);
		\path (1) edge[<-, draw=black]  (4);

		\path[draw,dashed,thick] (-0.4,3.7) rectangle (0.4,4.5);
		
		\path[draw,dashed,thick] (-0.4,1.6) rectangle (2.9,2.5);
		
		\path[draw,dashed,thick] (2.1,3.7) rectangle (2.9,4.5);

		\end{tikzpicture}  
		\caption{The preference-action model before the sale. The solid lines represent both $\leq_{i \to k}$ and $\leq_{j\to i}$. The dashed
			rectangles represent $j$'s relation $\sim_j$. }
		\label{tab:5}} 
\end{figure}

One model of the initial situation, before the sale, is depicted in
Figure~\ref{tab:5}. For simplicity we assume that $i$ and $k$
are passive in this case, so $\sim_i$ and $\sim_k$ is set to
$W \times W$. We consider three situations: :
\begin{itemize}
    \item $w_1$, in which the goods are in $k$'s possession and $i$ doesn't pay the 600 euro to her \emph{as a result of some action by $j$}. One can think of this case as one where $j$ sees to it that goods within the 500 euro budget get into $i$'s possession;
    \item $w_4$ in which the goods are in $i$'s possession and she pays the 600 euro to $k$, again as a result of some action by $j$;
    \item $w_2$ and $w_3$, which together represent the situation where $j$ abstains from doing
anything regarding $p$ and $f$. 
\end{itemize}
The facts that $k$ initially owns the
goods is represented by the
relation $\leq_{i \to k}$. It ranks $w_1$ highest, followed by $w_2$ and $w_3$
together, and $w_4$ lowest.  Since $i$ is $j$'s principal in this case, we assume as well that this legal ideality is the same for $j$ as far as $i$'s legal obligations are concerned, i.e. $\leq_{i \to j} = \leq_{j \to i} $. This gives, as desired,
that $j$ ought to see to it that she doesn't put $i$ in the position of having to pay the 600 euro to $k$: $O_{j \to i}Do_j\neg O_{i \to k} f$.

\begin{figure}[htbp]
	{\centering
		\begin{tikzpicture}[->,>=stealth',shorten >=1pt,shorten <=1pt,
		auto,node distance=.5cm,semithick,scale=.8]
		\tikzstyle{every state}=[draw=none,text=black]
		
		
		\node[state,label=left:{$Do_j \neg  p$}] at (0,1) (1)
		{{\small $a_1$}};
		\node[state,label=below:{$ \neg Do_j p \wedge \neg Do_j \neg p$}] at (4,1) (2)
		{{\small $a_2$}};
		\node[state,label=right:{$ Do_j p$}] at (8,1) (3)
		{{\small $a_3$}};

		\path (1) edge[|->, draw=black] (2);
		\path (2) edge[|->, draw=black] (3);
		\end{tikzpicture}
		
		\caption{The deontic action model $\mathcal{A}_j$ for $j$'s buying from $k$.}
		\label{tab:6}} 
\end{figure}
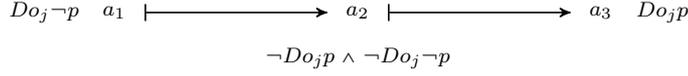

Agent $j$'s power to buy the 600 euro' worth of goods from $k$ is simply the power to transfer possession of those goods from $k$ to $i$, and creates a claim for the latter that the former pays the 600 euro. We model this as an action that results in making it the case that $i$ ought to be in possession of the goods and hence pays the 600 euro to $k$ . The deontic action
model of Figure~\ref{tab:6} has precisely that effect. 

\begin{figure}[htbp]
	{ \centering
		\begin{tikzpicture}[->,>=stealth',shorten >=.1pt,shorten <=.1pt,
		auto,node distance=.2cm,semithick, scale=.8]
		\tikzstyle{every state}=[draw=none,text=black]
		
		
		
		\node[state,label=left:{$\neg p$, $\neg f$}] at (0,5) (1)
		{{\small $(w_1, a_1)$}};
		\node[state,label=left:{$\neg p, \neg f$}] at (0,2) (2)
		{{\small $(w_2, a_2)$}};
		
		\node[state,label=right:{$p, f$}] at (3,5)
		(4) {{\small $(w_4, a_3)$}};
		\node[state,label=right:{$p, f$}] at (3,2)
		(3) {{\small $(w_3, a_2)$}};

		\path (1) edge[->, draw=black]  (2);
		\path (4) edge[<-, draw=black]  (3);
		\path (3) edge[<->, draw=black]  (2);
		\path (1) edge[->, draw=black]  (4);

		\path[draw,dashed,thick] (-.8,4.5) rectangle (0.8,5.5);
		
		\path[draw,dashed,thick] (-.8,1.5) rectangle (3.8,2.5);
		
		\path[draw,dashed,thick] (2.2,4.5) rectangle (3.8,5.5);

		\end{tikzpicture}  
		\caption{The preference-action model after the sale. The solid lines represent both $\leq_{i \to k}$ and $\leq_{j\to i}$, after the update. }
		\label{tab:7}} 
\end{figure}
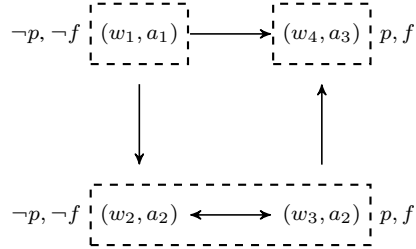

The result of updating the original model with the deontic action
model in Figure~\ref{tab:6} is represented in Figure~\ref{tab:7}. It
essentially reverts to the order of the strict preference relations $\leq_{i \to k}$ and $\leq_{j\to i}$,
leaving the indifference relation between $w_2$ and $w_3$ constant. This gives
$O_{i \to k} f$, as desired, as well as that $j$ ought to see to it that title to the property now goes to their new legitimate owner $i$:
$O_{j \to i}Do_j p$. Since none of those claim rights held before, we obtain as
desired that $j$ has indeed the (local) power to buy the goods from $k$: both
$\bigvee_{a \in A}\langle \A_j, a\rangle O_{i \to k}f$ and
$\bigvee_{a \in A}\langle \A_j, a\rangle O_{j \to i}Do_j p$ hold in the
original model.

\subsection{Impermissibility of Buying} \label{sec:legperm} 

The language $\mathcal{L}$, even extended with the dynamic modalities,
cannot directly express the notion of legal permissibility of deontic
actions. This language is designed to describe the effects of deontic
action or, by having deontic modalities scoping over dynamic
expressions, the normative status of those effects. But this is still
different from saying that a certain deontic action is obligatory,
permitted or forbidden. In our example it is arguably the case that it
ought to be that $i$ owns the goods $j$ has bought in her name, even
though this transaction constitutes a violation of the contract's conditions.

Our modeling of the legal permissibility of deontic action is inspired by
the Anderson-Kanger reduction of deontic modalities to a combination
of alethic ones with a ``violation'' or ``sanction'' constant,
c.f.~\citep{kanger1970new,meyer1988different}. Let $V$ be that
constant, and $a$ one of $i$'s deontic actions, then this reduction
goes as follows:
$$P(a) \equiv_{df} \langle \A_i, a\rangle \neg V  $$
Impermissible deontic actions can be straightforwardly represented
using the post-condition function, since there being a violation is
itself a normative, or at least an evaluative notion, which might
be changed by the execution of certain deontic actions. A higher court
can, for instance, invalidate a guilty verdict from a lower one,
effectively changing whether a violation has occurred or not. In our
example, for instance, we can simply set $$Post(a_i)(V) := \top $$ for
all $i = 1, 2, 3$. This immediately leads to the result that buying the goods from $k$ is not legally permissible, i.e.
$\bigwedge_{a \in A}\neg P(a)$ will be true in all states in the
original model. By setting $V$ false everywhere before the update we
furthermore obtain the welcome consequence that, before the sale, $j$ is
not in a state of violation, in line with the fact that $O_{j \to i}Do_jf $ is
false in all states before the update.

\section{Conclusion}
This paper has studied the potential of a deontic re-interpretation of a formalism
developed in dynamic epistemic logic to model Hohfeldian legal competences. We have shown that it allows for a model of Hohfeldian power and immunity that, unlike
Lindahl's~\citeyearpar{lindahl1977position} approach, explicitly
captures the norm-changing or dynamic character of legal
competences. It does so both at the semantic level, through the
explicit update mechanism, and at the syntactic level, by using an
explicit dynamic modality to express the effects of deontic
actions. The approach we propose here is, however, also reductive in
the sense that formulas with dynamic modalities are semantically and
provably reducible to formulas without. As a result, it comes with a
rich set of interaction principles between static and dynamic
rights, captured by the so-called "reduction axioms." 
We have pointed out in Section 2, however, that in view of its complexity, the reduction of dynamic rights to static language only yields limited insights on the relation between both levels. 
Finally, we have shown that this
system can capture the distinction between legal ability and legal
permissibility in a more auspicious way than reductive approaches,
without paying the price of thorough-going non-reductionism.

We take this to be a starting point for the methodology we
propose, but of course it also raises a number of questions that could
not be addressed in this paper. The natural next step is to study the
theory of normative positions stemming from our models of dynamic
rights. See the thesis~\citep{dong2017thesis} for some steps in that direction.
Equally important in our view is to study the theory of legal
competences that would result from extending a static base that is
different from standard deontic logic. In the epistemic context a wide
variety of static logics of knowledge and belief have been
``dynamified'' using the action model methodology. More radical
departures from standard deontic logic have been proposed to capture actual legal
reasoning, and the question remains whether they would yield a
plausible theory of power and immunity once augmented with a dynamic
module as we have done here. Finally, there is of course an interesting connection between legal competences as we have represented them and a dynamic analysis of the notion of strong permission~\citep{vonWright1963}, as for instance suggested by~\cite{benthem2014priority}. This, however, is the topic of another paper.




\bibliographystyle{spbasic}
\bibliography{bib}

\end{document}